\documentclass[12pt]{article}
\usepackage{epsfig}
\usepackage{amssymb,graphicx}
\def\be{\begin{equation}}
\def\ee{\end{equation}}
\def\ba{\begin{array}{c}}
\def\ea{\end{array}}

\newcommand{\bea}{\begin{eqnarray}}
\newcommand{\eea}{\end{eqnarray}}

\newcommand{\kt}{\rangle}

\begin{document}

\begin{center}

{\Large \bf {

Quantum square well with logarithmic central spike

 }}

\vspace{13mm}

\vspace{3mm}

\begin{center}

\textbf{Miloslav Znojil}\footnote{znojil@ujf.cas.cz}
 and
\textbf{Iveta Semor\'{a}dov\'{a}}\footnote{semoradova@ujf.cas.cz}

\vspace{1mm} Nuclear Physics Institute of the CAS, Hlavn\'{\i} 130,
250 68 \v{R}e\v{z}, Czech Republic

\end{center}

\vspace{3mm}

\end{center}

\subsection*{Keywords:}

.

state-dependence of interactions;

effective Hamiltonians;

logarithmic nonlinearities;

linearized quantum toy model;

\subsection*{PACS number:}
.

PACS 03.65.Ge – Solutions of wave equations: bound states

\section*{Abstract}

Singular repulsive barrier $ V(x)= -g \ln (|x|)$ inside a square
well is interpreted and studied as a linear analogue of the
state-dependent interaction ${\cal L}_{eff}(x) =
-g\,\ln[\psi^*(x)\psi(x)]$ in nonlinear Schr\"{o}dinger equation. In
the linearized case, Rayleigh-Schr\"{o}dinger perturbation theory is
shown to provide a closed-form spectrum at the sufficiently small
$g$ or after an amendment of the unperturbed Hamiltonian. At any
spike-strength $g$, the model remains solvable numerically, by the
matching of wave functions. Analytically, the singularity is shown
regularized via the change of variables $x=\exp y\ $ which
interchanges the roles of the asymptotic and central boundary
conditions.

\newpage

\section{Motivation\label{Ia}}

The study of complicated quantum systems may be facilitated by a
judicious, less explicit treatment of certain less essential degrees
of freedom \cite{Feshbach}. The reduction may lead to
simplifications which are often achieved via a tentative replacement
of the exact, full-Hilbert-space linear Schr\"{o}dinger equation
 \be
 {\rm i}\partial_t \,|\Psi\kt = \mathfrak{H}\,|\Psi\kt
  \label{seor}
 \ee
by its reduced, open-subsystem version. In this manner a remarkably
successful description of the physical reality has been achieved, in
multiple phenomenological applications (cf., e.g., their compact
review in paper I \cite{Kostya}) via nonlinear evolution equations
in which the effective-interaction potential was admitted
state-dependent,
 \be
  {\rm i} \partial_t \psi({x},t)
  =\left [- \partial_x^2 +V_{(eff)}({x},t)\right ]\, \psi({x},t)\,, \ \ \ \
  \ \ \ \ V_{(eff)}({x},t) = V[\psi({x},t),x,t]
  \label{ser}
  \ee
and, in particular, in which the nonlinearity has been chosen
logarithmic,
 \be
 V[\psi({x},t),x,t] = -g\,\ln[\psi^*(x,t)\psi(x,t)]\,,
 \ \ \ \ g > 0
 \,.
 \label{nelie}
 \ee
In paper I it has been argued that an important formal support of
the latter choice may be seen in its asymptotic system-confinement
self-consistency. Indeed, one can easily verify that the insertion
of a tentative, exactly solvable harmonic-oscillator potential
$V_{(HO)}(x)=x^2$ in Eq.~(\ref{ser}) would yield the wave functions
in closed form
 \be
 \psi_{(HO)}({x},t) \sim \exp(-i E_{(HO)} t)\,
\exp\left(-{x}^2/2
 +{\cal O}(\ln|{x}|)\right)\,,
 \ \ \ \ \ \ \ \ |x| \gg 1\,.
 \label{HOsample}
 \ee
In turn, expression (\ref{nelie}) will lead to the qualitatively
correct asymptotics
 \be
 -\ln[\psi^*_{(HO)}({x},t) \psi_{(HO)}({x},t) ] =
  {x}^2
 +{\cal O}(\ln{|{x}|})\,,
 \ \ \ \ \ \ \ \ |x| \gg 1\,
 \label{rela}
 \ee
of the confining potential.

In the constructive part of paper I it has been recalled and
demonstrated that certain node-less, ``gausson'' solutions of the
evolution Eq.~(\ref{ser}) may be well-behaved at all times $t$. The
existence of the ``gaussons'' can be viewed as a consequence of the
absence of the nodal zeros in the initial (i.e., say, $t=0$) choice
of the ground-state-like wave function $\psi(x,0)$. In other words,
what remained unclarified in paper I was the question of the
properties of all of the non-gausson solutions of Eq.~(\ref{ser}) +
(\ref{nelie}). Presumably, most of these non-gausson solutions might
be based on the anomalous initial wave functions $\psi_N(x,0)$
having an $N-$plet of the excited-state-like nodal zeros at some $N
\geq 1$.

Naturally, this would make the nonlinear interaction term
(\ref{nelie}) singular. Locally (i.e., out of the asymptotic region
and near any nodal zero $x_j$ with $1 \leq j \leq N$) this follows
from the simple-zero estimate
 \be
 \psi_N(x,0)\sim x-x_j\,.
 \label{six}
 \ee
In the first nontrivial non-gausson case we may choose $N=1$ and
require $x_1=0$ (i.e., an initial-time antisymmetry of the wave
function). In the light of the above-mentioned HO example one
expects that, with certain implicit, not too well tractable error
terms, we should work with potentials of the form
 \be
 V_{(eff)}({x},0) \ \sim \ x^2 -2g \ln |x| + \ldots\,
 \ee
or, in an alternative, technically simpler confining square-well
(SW) approximation, with
 \be
 V_{(eff)}({x},0) \ \sim \ V_{(SW)}({x}) =
 \left \{
 \begin{array}{cl}
 \infty\,, \ \ \ & |x|\geq 1\,,\\
  -2g \ln |x|\,, \  & |x| < 1\,.
  \ea
  \right .
  \label{linearized}
 \ee
In what follows we intend to complement the global non-linear-theory
considerations of paper I by the missing and relevant discussion of
some of the technical consequences of the emergence of the
logarithmic singularity at one or more nodal points $x_j$ and, first
of all, of certain properties of the bound states in the first
nontrivial excitation-simulating {\em linear} interaction potential
(\ref{linearized}).

\section{Weak-coupling regime\label{perto}}

The enormous simplicity of the one-dimensional quantum square well
oscillator makes it suitable for pedagogical purposes. Its analyses
appear not only in conventional textbooks \cite{Fluegge,Flueggeb}
but also in the less conventional studies of supersymmetric quantum
systems \cite{susypt,susyptb}. The elementary nature of the
square-well model found also nontrivial applications in parity times
time-reversal symmetric quantum mechanics
\cite{ptsqw,ptsqwb,fragile,fragileb} or in certain sophisticated
versions of perturbation theory \cite{Langer}.

%

\begin{figure}[h]                    
\begin{center}                         
\epsfig{file=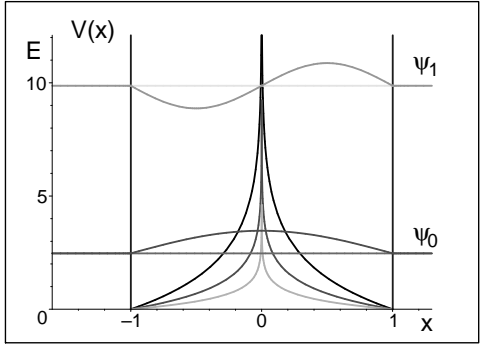,angle=270,width=0.36\textwidth}
\end{center}    
\vspace{2mm} \caption{The shapes of the perturbed square-well potentials (\ref{linearized}) at
$g=1/4$ (the lowest, light spike), $g=1/2$ (the intermediate spike) and
$g=1$ (the upper spike). Horizontal lines mark the first two
energy levels for $g=0$. Attached to them, the picture also displays the shapes of the unperturbed
wave functions.
 \label{trija}
 }
\end{figure}

We intend to analyze the interaction in its special form (\ref{ser})
+ (\ref{linearized}), i.e., in the first nontrivial $N=1$ special
case. Thus, our attention gets shifted to the linearized model
represented by the conventional and time-independent ordinary
differential Schr\"{o}dinger equation for quantum stationary bound
states,
 \be
 -\, \frac{{\rm d}^2}{{\rm d} x^2}\, \psi_n(x)
 -2g\,\ln (|x|)\, \psi_n(x)= E_n\,
 \psi_n(x)\,,
 \ \ \ \ \ \
 \psi_n(\pm 1) =0\,,
 \ \ \ \ n = 0, 1, \ldots
 \,.
   \label{SEx}
  \ee
The spike-shaped logarithmic barrier is unbounded, repulsive and
centrally symmetric here - cf. Fig.~\ref{trija} where the shape of
the potential is displayed at $g=0.25, 0.5$ and $1$.

It is worth adding that although we are choosing here (i.e., in
Eq.~(\ref{six})) the first nontrivial nodal number $N=1$ for the
sake of simplicity, Eq.~(\ref{SEx}) describes {\em all} of the bound
states generated by the linearized toy-model interaction
(\ref{linearized}). These states are numbered by a {\em different},
lower-case index $n$. Naturally, one would have to set, for the sake
of consistency, $n=N$ at the end of the analysis and, in principle
at least, before an ultimate return to the initial
nonlinear-equation setting.


\subsection{First-order approximation}

At the small but non-vanishing strengths $g>0$ our central barrier
is infinitely high so that one might suspect that it is
impenetrable. The expectation is wrong. In the
Rayleigh-Schr\"{o}dinger perturbation-expansion series arrangement
\cite{Messiah} with
 \be
 E_n=E_n(g)=E_n^{[0]} + g\,E_n^{[1]}+ g^2\,E_n^{[2]}  + \ldots\,,
 \ \ \ \ \ \ E_n^{[0]}= [(n+1)\pi/2]^2\,,
 \ \ \ \ n = 0, 1, \ldots
 \label{ten}
 \ee
the first-order shifts
 \be
 E_{2p}^{[1]} \sim - 4\,\int_{0}^{1} \cos^2 [(2p+1) \pi\, x/2]\,\ln
 (|x|)\,dx\,,\ \ \ \ \ \ p = 0, 1, \ldots
 \label{suda}
 \ee
and
 \be
 E_{2q+1}^{[1]} \sim - 4\,\int_{0}^{1} \sin^2 [(q+1) \pi\, x]\,\ln
 (|x|)\,dx\,,\ \ \ \ \ \ q = 0, 1, \ldots
 \label{licha}
 \ee
of the spectrum of our conventional perturbed square well are,
obviously, positive and finite.

One can easily prove that the latter corrections are all finite,
indeed. The proof may be based on the fact that all of the
integrands are bounded on the subintervals of $x \in (\epsilon,1)$
with any $\epsilon = \exp(-R) \in (0,1)$ (i.e., any $R>0$). Thus, it
is sufficient to find a bound for the integrals over a short
interval of $x \in (0,\epsilon)$. With a sufficiently large $R \gg
1$ we obtain an explicit estimate
 \be
 - 4\,\int_{0}^{\epsilon} \sin^2 [(q+1) \pi\,x]\,\ln
 (|x|)\,dx
 \sim \int_{0}^{\epsilon}\,x^2\,\ln
 (|x|)\,dx =\int_{-\infty}^{-R}\,y\,e^{3y}
 \,dy= -\frac{1}{9}(3R+1)\,e^{-3R}\,
 \label{alicha}
 \ee
showing that the odd-state integrals are all exponentially small. In
the even-state case we have, similarly, the estimate
 \be
 - 4\,\int_{0}^{\epsilon} \cos^2 [(2p+1) \pi\,x/2]\,\ln
 (|x|)\,dx
 \sim \int_{0}^{\epsilon}\,\ln
 (|x|)\,dx =\int_{-\infty}^{-R}\,y\,e^{y}
 \,dy= -(R+1)\,e^{-R}\,
 \label{asuda}
 \ee
leading to the same ultimate finite-correction conclusion.

\subsection{Closed formulae}

The even-state contribution seems larger than the odd-state
contribution, for the sufficiently large $R$ at least. For a more
reliable, $R-$independent comparison of the corrections
$E_{n}^{[1]}$ at the even and odd $n$ it is necessary to introduce
the conventional sine-integral special functions $Si(x) = \int_0^x\,
\sin(t)/t \,dt\ $ and to evaluate the first-order corrections
exactly. Fortunately, this is feasible yielding the following
formulae,
 \be
 E_{2p}^{[1]} = 2 + \frac{2}{(2p+1)\pi}\,Si [(2p+1)\pi]\,,
 \ \ \ \ p = 0, 1, \ldots\,,
 \ee
 \be
 E_{2q+1}^{[1]} = 2 - \frac{2}{(2q+2)\pi}\,Si [(2q+2)\pi]\,,
 \ \ \ \ q = 0, 1, \ldots\,.
 \ee

\begin{table}[h]
\caption{The first ten numerical coefficients $E_{n}^{[1]}$ in
Eq.~(\ref{ten}). } \label{unota}

\vspace{2mm}

\centering
\begin{tabular}{ccc}
\hline \hline
   $ n$ & symmetric $\psi_n(x)$ & antisymmetric  $\psi_n(x)$ \\
 \hline
 \hline
   0 & 3.178979744 &\\
   1&&  1.548588333\\
   2 & 2.355395491 &\\
   3 && 1.762515165 \\
  4 &  2.208042866&\\
   5 &&  1.838931594 \\
   6 & 2.146975999 &\\
   7 && 1.878156443\\
   8 & 2.113606700 &\\
   9 && 1.902022366 \\
 \hline
 \hline
\end{tabular}
\end{table}

The evaluation of the numerical values of the special functions
$Si(x)$ is routine and yields the first-order perturbation-series
coefficients as sampled in Table \ref{unota}. The inspection of the
Table reveals that the energy shifts at the even quantum numbers
$n=2p$ will be always larger than the partner shifts at the odd
quantum numbers $n=2p+1$. In other words, even the present ``soft'',
logarithmic shape of the central repulsive barrier will lead to the
quasi-degeneracy pattern in the spectrum of the logarithmically
spiked bound states.

\subsection{Limitations of applicability}

By the logarithmically singular but still positive-definite barrier
the unperturbed spectrum is being pushed upwards. Still, due to the
immanent weakness of the singularity of the logarithmic type even in
the strong-coupling dynamical regime with $g \gg 1$, the expected
effect of the quasi-degeneracy will get quickly suppressed with the
growth of $n$, i.e., of the excitation. In contrast to the stronger
and more common (e.g., power-law) models of the repulsion in the
origin, this will make the real influence of the logarithmic barrier
restricted to the low-lying spectrum.

Fig.~\ref{hoja} may be recalled for an explicit quantitative
illustration of the latter expectation. Using just the most
elementary leading-order-approximation estimates we see there that
while the prediction of the quasi-degeneracy between the ground
(i.e., $n=0$) and the first excited (i.e., $n=1$) state might still
occur near the reasonably small value of coupling $g_{0,1}^{[1]}=
4.540138798$, the next analogous crossing of the first-order
energies $E_2^{[1]}$ and $E_3^{[1]}$ only takes place near the
estimate as large as $g_{2,3}^{[1]}=29.13203044$, etc.

%

\begin{figure}[h]                    
\begin{center}                         
\epsfig{file=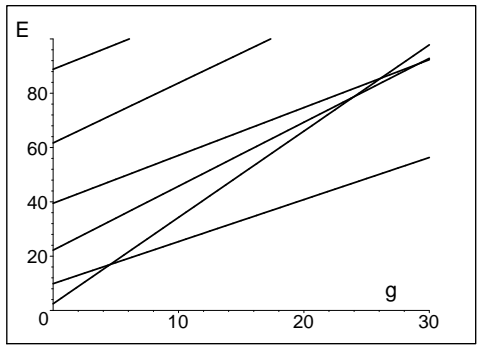,angle=270,width=0.36\textwidth}
\end{center}    
\vspace{2mm} \caption{The coupling-dependence of the low-lying spectrum in the
Rayleigh-Schr\"{o}dinger
first order
(i.e., linear-extrapolation) approximation.
 \label{hoja}
 }
\end{figure}

The same Fig.~\ref{hoja} also shows that another first-order
crossing may be detected for $E_0^{[1]}$ and $E_2^{[1]}$, emerging
even earlier (i.e., at $g_{0,2}^{[1]}=23.96744320$) and being,
obviously, spurious. In other words, for the prediction of the
quasidegeneracy in the strong-coupling dynamical regime the
knowledge of the mere first-order perturbation
corrections must be declared
insufficient.

\section{Strong-coupling regime}

Beyond the domain of applicability of the weak-coupling perturbation
theory, alternative (mainly, purely numerical) methods must be used
in order to determine the spectrum of bound states of our linearized
toy model exactly, i.e., with arbitrary prescribed precision.

As long as these bound states are determined by the ordinary
differential Schr\"{o}dinger Eq.~(\ref{SEx}), there exists a number
of methods of their construction. The choice of the method may be
inspired by the inspection of Figs.~\ref{trija} and \ref{hoja}. This
indicates that the influence of our singular logarithmic potential
(\ref{linearized}) is felt, first of all, by the low-lying bound
states and/or in the strong-coupling regime. Directly, this may be
demonstrated by the routine numerical construction of the
wavefunctions (sampled in Fig.~\ref{fceja}) and by the routine
numerical evaluation of the energies (sampled in
Table~\ref{triota}). In both cases, due attention must be paid to
the singular nature of our potential (\ref{linearized}) in the
origin. This is a challenging aspect of the numerical calculations
which will be discussed and illustrated by some examples in what
follows.


%
%
%
%

\begin{figure}[h]                    
\begin{center}                         
\epsfig{file=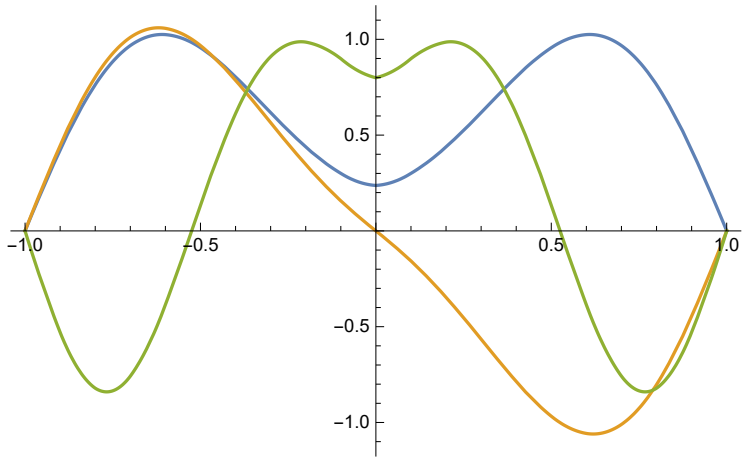,angle=0,width=0.56\textwidth}
\end{center}    
\vspace{2mm} \caption{The first three wavefunctions at $g=10$.
 \label{fceja}
 }
\end{figure}

\begin{table}[h]
\caption{Numerically determined low-lying spectrum. } \label{triota}

\vspace{2mm}

\centering
\begin{tabular}{cccccc}
\hline \hline
   $g$ &  $E_0$  &  $E_1$ &  $E_2$ &  $E_3$ &  $E_4$\\
 \hline
 \hline
0.00&                            2.4674
   &                          9.8696
    &                         22.207
     &                       39.478
      &                       61.685\\
0.25 &3.2478 & 10.255 & 22.796 & 39.928 & 62.265\\
0.50 & 4.0097 & 10.638 & 23.394 & 40.369 & 62.817 \\
1.00 & 5.4784 & 11.395 & 24.618 & 41.252 & 63.931\\
    \hline
 \hline
\end{tabular}
\end{table}

\subsection{Mathematical challenge: singularity}

Without any real loss of generality our attention may stay
restricted to the characteristic $N=1$ model (\ref{SEx}). This model
is sufficiently general when treated as a system of two (viz., $x<0$
and $x>0$) linear differential equations for which the logarithmic
derivatives of the respective wave functions have to be matched in
the origin, i.e., at $x_1=0$. In this sense, naturally, the
generalization to the $N>1$ cases would be straightforward.

Once we fix $N=1$ we may split the original linear differential
equation into a  pair living on the two respective half-intervals,
viz.,
 \be
 -\, \frac{{\rm d}^2}{{\rm d} x_{(L)}^2}\, \psi_n(x_{(L)})
 -2g\,\ln (-x_{(L)})\, \psi_n(x_{(L)})= E_n\,
 \psi_n(x_{(L)})\,,
 \ \ \ \ \ \ x_{(L)} \in (-1,0)\,
   \label{SExL}
  \ee
and
 \be
 -\, \frac{{\rm d}^2}{{\rm d} x_{(R)}^2}\, \psi_n(x_{(R)})
 -2g\,\ln (x_{(R)})\, \psi_n(x_{(R)})= E_n\,
 \psi_n(x_{(R)})\,,
 \ \ \ \ \ \ x_{(R)} \in (0,1)\,.
   \label{SExR}
  \ee
As long as the bound-state solutions have a definite parity at
$N=1$, one of these equations may be omitted as redundant.

\subsection{Amended zero-order approximation\label{pese}}

Due to the symmetry of our potential it will be sufficient to
consider just Eq.~(\ref{SExL}) on the negative finite half-interval.
The correct matching with Eq.~(\ref{SExR}) may be then guaranteed by
means of the pair of {\em ad hoc\,} boundary conditions in the
origin,
 \be
   \left \{
 \begin{array}{ll}
  \psi_n(0) \neq 0
 \,,\  \  \psi_n'(0) = 0
 \,,\  \ & {\rm (even\ states)}\,,\\
  \psi_n(0) = 0
 \,,\  \  \psi_n'(0) \neq 0
 \,,\  \
   & {\rm (odd\ states)}\,.\\
 \ea
 \right .
 \,
   \label{hox}
  \ee
The application of recipe (\ref{hox}) must be performed with due
care. The reason lies in the unbounded, singular nature of our
logarithmic potential in the origin. Fortunately, the necessary
deeper analysis of the matching conditions at $x=0$ might be also
complemented by an amendment of the perturbation-theory
considerations of section~\ref{perto}.

Inside a suitable small interval of $x  \in (-d,d)$ with $d<1$ we
may contemplate an approximative replacement of the logarithmic
repulsive spike $V(x)$ by a rectangular barrier of a finite height
$\kappa_0^2= E + \kappa^2$. At any given/tentative energy $E$ we may
demand that $E=k^2=-2g\ln d$, i.e., that $d = d(E) = \exp
[-E/(2g)]$. This will weaken our potential $V()x)$ to the left from
$x=-d$ while strengthening it, locally at least, to the right from
$x=-d$. The approximate wavefunctions may be then written down in
the following closed elementary-function form
 \be
  \psi(x_{})
 \approx
 \left \{
 \begin{array}{lll}
 \sin [(x+1) k]\,,\  \ & x_{} \in (-1,-d)\,,&\\
   \sin [(x-1)k ]\,,\  \ & x_{} \in  (d,1)\,,&\\
  {\rm cosh} [\kappa x ]\,,\  \ & x_{} \in (-d,d)\, \ & {\rm (even\ states)}\,,\\
  {\rm sinh} [\kappa x ]\,,\  \ & x_{} \in (-d,d)\, \ & {\rm (odd\ states)}\,.\\
 \ea
 \right .
 \,
   \label{oxLav}
  \ee
Its components must be properly matched at $x=\pm d=\pm d(E)$ and,
if needed, also properly normalized. It is, perhaps, worth adding
that inside the interval of $E\in (5,5.5)$ (i.e., in the vicinity of
the gound-state energy, cf. Table \ref{triota}), the decrease of the
value of $d(E)$ from
                           $d(5)\approx 0.08208499862$
                      to   $d(5.5) \approx   0.06392786121$
                      is almost linear, i.e.,
                      comparatively easily kept under control
                      in the calculations.
This means that the $d-$dependence of our approximative rectangular
barrier may be also re-interpreted as its energy-dependence.

\subsection{Regularized Schr\"{o}dinger equation}

The remarkable technical challenge is that the point of the matching
of the wavefunctions coincides with the singularity of the
potential. In the literature, similar situation is usually
encountered in connection with the Coulombic repulsion $\sim 1/x$
\cite{Coulombic,Coulombicb}. Here, the repulsion is much weaker but
the regularization of the model is still by far not routine. In the
analytic-function-theory language it can rely upon the change of
variables
 \be
 x_{(L)}=-\exp(-\lambda)\,,
 \ \ \ \ \ \ \
 x_{(R)}=\exp(\rho)\,,
 \label{19}
 \ee
and
 \be
 \psi_n(x_{(L)})=\exp(-\lambda/2)\,\phi^{(-)}(\lambda)\,,
 \ \ \ \ \ \ \
 \psi_n(x_{(R)})=\exp(\rho/2)\,\phi^{(+)}(\rho)\,.
 \label{20}
 %
 \ee
In the left subinterval of $x_{(L)} \in (-1,0)$ this yields the
initial-value problem
 \be
 -\, \frac{{\rm d}^2}{{\rm d} \lambda^2}\, \phi(\lambda)
 +\frac{1}{4}\, \phi(\lambda)=\exp(-2\lambda)\,
 \left [E-2g\lambda
 \right ]\, \phi(\lambda),%
 \ \ \ \
 \phi(0) =0\,,
 \ \ \ \
 \phi'(0) =1\,
   \label{ExLm}
  \ee
on the half-line of $\lambda \in (0,\infty)$.

We see that the singularity has been moved to infinity. Thus, after
the change of variables (\ref{20}), the wavefunction-matching
relations (\ref{hox}) may purely formally be replaced by their
appropriate asymptotic analogues supplementing Eq.~(\ref{ExLm}). In
fact, we shall show below that this is really a purely analytic
idea, not leading to any practical numerical advantages.

Analogously, the second half of our initial differential equation
with $x_{(R)} \in (0,1)$ is converted into equivalent second
half-line version
 \be
 -\, \frac{{\rm d}^2}{{\rm d} \rho^2}\, \phi(\rho)
 +\frac{1}{4}\, \phi(\rho)=\exp(2\rho)\,
 \left [E+2g\rho
 \right ]\, \phi(\rho),
 \ \ \ \
 \phi(0) =0\,,\ \ \ \
 \phi'(0)= (-1)^{n+1}=\pm 1\,
   \label{ExRm}
  \ee
where $\rho \in (-\infty,0)$. The latter equation is redundant
again, obtainable from the former one by the mere change of parity
{\it alias\,} replacement $\rho \to -\lambda$.

\section{Numerical solutions}

Using our amended approximation (\ref{oxLav}) we could obtain a {\em
qualitatively\,} correct shape of the wavefunctions, in principle at
least. Naturally, the fully reliable construction of bound states
must be performed by the controlled-precision numerical integration
of our ordinary differential Schr\"{o}dinger equations. These
results were sampled in Fig.~\ref{fceja} above. What is worth
emphasizing is that in this setting the singularity of potential
$V(x)$ remains tractable by the standard numerical-integration
software, say, of MATHEMATICA or MAPLE.

\subsection{Qualitative theory: re-parametrized Schr\"{o}dinger equation}

For our model (\ref{SEx}), obviously, an extension of applicability
of the conventional perturbation theory could have been based on the
use of various more sophisticated zero-order shapes of $V_0(x)$. In
particular, we could obtain a more quickly convergent sequence of
perturbation approximations when using the specific
rectangular-potential choice of $V_0^{(RP)}(x)$ of paragraph
\ref{pese}.

One of the other benefits of the amendment would be qualitative,
based on the observation that a small deformation of the special
potential $V_0^{(RP)}(x)$ of paragraph \ref{pese} must lead just to
a small deformation of the related trigonometric definitions
(\ref{oxLav}) of the wavefunctions.

Via these deformations, we may even return {\em exactly} to our
full-fledged logarithmic potential $V(x)$. Such a return would be
achieved by means of the replacement of the effective-momentum
constant $k=\sqrt{E}$ by a weakly coordinate-dependent function
$\mu(E,x)=\sqrt{E-V(x)}$. The other constant $\kappa=\kappa(E)$ gets
replaced by the effective barrier-height $\nu(E,x) =\sqrt{V(x)-E}$.
As a result, this enables us to rewrite our Schr\"{o}dinger equation
in the partitioned form
 \be
 -\, \frac{{\rm d}^2}{{\rm d} x_{}^2}\, \psi(x_{})
 =
 \left \{
 \begin{array}{ll}
 \ \ \mu^2(E,x)\, \psi(x_{})\,,\  \ & x_{} \in (-1,-d)\bigcup (d,1)\,,\\
 -\nu^2(E,x)\, \psi(x_{})\,,& x_{} \in (-d,d)\,.\\
 \ea
 \right .
 \,
   \label{SExLav}
  \ee
By construction, this equation is exact. Still, in the light of
Eq.~(\ref{oxLav}) it may be assigned the {\em approximate}
elementary solutions
 \be
  \psi(x_{})
 \approx \pm\,
 \left \{
 \begin{array}{lll}
 \sin [\mu(E,x)(x+1) ]\,,\  \ & x_{} \in (-1,-d)\,,&\\
  \sin [\mu(E,x)(x-1) ]\,,\  \ & x_{} \in  (d,1)\,,&\\
  {\rm cosh} [\nu(E,x)x ]\,,\  \ & x_{} \in (-d,d)\, \ & {\rm (even\ states)}\,,\\
  {\rm sinh} [\nu(E,x)x ]\,,\  \ & x_{} \in (-d,d)\, \ & {\rm (odd\ states)}\,.\\
 \ea
 \right .
 \,
   \label{hoxLav}
  \ee
Once we take into account the parity, the matching in the origin
remains trivial. The decisive role will now be played, instead, by
the smoothness (i.e., sort of matching) of the amended approximate
wave functions (\ref{hoxLav}) at the energy-dependent point $x=-d<0$
or, equivalently, at $x=d>0$.

\subsection{The danger of ill-conditioning}

The wavefunction $\psi_0(x)=\psi(x)$ is nodeless and spatially
symmetric. It must obey the ordinary differential Eq.~(\ref{SExL})
on half-interval so that its construction may proceed numerically.
This yields the $x \to 0^-$ limiting values which must be made
compatible with the upper line of Eq.~(\ref{hox}) via a suitable
choice of energy $E$. Such an algorithm the leads to the results
sampled by Table~\ref{triota}.

In principle, we could also employ the change of variables
(\ref{19}) and (\ref{20}) and replace Eq.~(\ref{SExL}) by its
equivalent version (\ref{ExLm}). This enables us to transfer the
central left-right matching (\ref{hox}) of wavefunctions to its
analogue in infinity, i.e., in the limit $\lambda \to +\infty$.

The key merit of such an arrangement lies in the clear picture of
the analyticity properties and, in particular, in the asymptotic
negligibility of the right-hand side of Eq.~(\ref{ExLm}). For this
reason the new equation may be immediately assigned the elementary
general asymptotic solution
 \be
 \phi(\lambda,E)= const\, e^{\lambda/2+ {\rm corrections}}
  + f(E) e^{-\lambda/2+ {\rm corrections}}\,.
 \ \ \ \ \ \
 \lambda \gg 1\,.
 \label{error}
 \ee
The changes of the energy only influence here the subdominant term
which is {exponentially small}. This means that {\em any\,}
numerical solution of the initial-value problem (\ref{ExLm}) will
remain insensitive to the variations of the tentative energy. Hence,
the task is ill-conditioned.

\begin{table}[h]
\caption{The failure of the backward-iteration method based on
Eqs.~(\ref{ExLm}) and (\ref{28}). At $q=1$, the convergence of
$\phi(0)\to 0$ with respect to $\lambda_{\rm max}\to \infty$ is too
slow. } \label{duota}

\vspace{2mm}

\centering
\begin{tabular}{cccc}
\hline \hline
   $E$ &  $\lambda_{\rm max} $ & evaluated $\phi(0)$ & difference \\
 \hline
 \hline
   5.55& 3.50& -0.064333935& -\\
  & 3.75&  -0.059104634& -0.0052\\
 & 4.00&  -0.053830824& -0.0053\\
   & 4.25& -0.048788380& -0.0050\\
   \hline
   5.45& 3.50& -0.037417250& - \\
  & 3.75&  -0.031754144& -0.0057\\
 & 4.00&  -0.026113710& -0.0056 \\
   & 4.25& -0.020763014& -0.0053 \\
     \hline
 \hline
\end{tabular}
\end{table}

One has to try to start the reconstruction in the opposite direction
initiated at a sufficiently large $\lambda_{\rm max} \gg 1$ via a
suitable tentative choice of the initial values of the wavefunction
$\phi(\lambda)$ and of its first derivative.

In a test study of the ground state at $g=1$ we made use of our
knowledge of the first-order perturbation result $E \approx
(\pi/2)^2+3.178979744=
                             5.646380845$
of section~\ref{perto}. Unfortunately, {\em any} choice of the
asymptotic initial values of wavefunctions seems to be destabilized
by certain uncontrolled numerical rounding errors as well. This is
an empirical observation sampled in Table~\ref{duota} in which we
employed the simplest possible choice of the tentative asymptotic
initialization
 \be
 \phi(\lambda_{\rm max})=1\,,
 \ \ \ \ \ \phi'(\lambda_{\rm max})=0\,.
 \label{28}
 \ee
The inspection of the Table reconfirms the scepticism evoked by the
analytic formula for wavefunction asymptotics  (\ref{error}) which
are numerically ill-conditioned.

\section{Discussion}


In the nonlinear Schr\"{o}dinger equation context as formulated and
reviewed in paper I the restriction of the constructive attention to
the mere node-less gaussons (i.e., to $N=0$) really weakened the
authors' original intention of making the asymptotically confining
interaction (\ref{nelie}) truly excitation-dependent (i.e., more
precisely, number-of-nodal-zeros-dependent).

On the basis of results of the preceding section we may now
conjecture that in practice the true impact of the presence of the
nodal zeros will be probably much smaller than expected. Although
these zeros induce the infinitely high barriers in the nonlinear
effective potentials (\ref{nelie}), these barriers remain penetrable
and narrow.

We saw here that in the context of perturbation theory the latter
properties of the logarithmic barriers render the quantitative
considerations feasible. The surprising, not entirely expected
friendliness of the perturbation analysis of toy model (\ref{SEx})
encourages also the use of the other, more universal numerical
methods.

\subsection{Linear models with $N>1$\label{sII}}

One of our key results may be seen in the observation that in the
technical sense one need not feel afraid of the presence of the
logarithmic (i.e., as we demonstrated, weak) singularities in the
interaction potentials, linear or not. In particular, in the linear
case one may feel encouraged to employ the standard techniques of
the matching of the piecewise analytic wave functions at the nodal
points $x_j$. In this context the readers may be recommended to have
a look at a few recent constructive analyses \cite{Ryu,Ryub} of the
similar scenarios.

In our present note we skipped the concrete numerical implementation
of the matching recipe. We have only pointed out that due to the
logarithmic nature of the singularity of the potential
(\ref{linearized}), one has to keep in mind that the most natural
change of variables (\ref{19}) + (\ref{20}) transforms the origin of
$x_{(L,R)}$ into infinities of $\lambda$ and $\rho$, and vice versa.
Still, we believe that this would cause just a minor complication in
numerical setting, more than compensated by the simplification of
the differential equation.

After the change of the variables, all of the basic features of the
conventional matching method remain unchanged. As long as in the new
setting of Eqs.~(\ref{ExLm}) and/or (\ref{ExRm}) the
energy-representing parameter $E$ becomes multiplied by an
exponential function, one should speak, strictly speaking, about the
so called Sturmian eigenvalue problem.

\subsection{Towards the nonlinearities\label{II}}

In the majority of the phenomenological scenarios, the predictions
provided by the guess of the {\em linear} interaction $V(\vec{x})$
may fail to fit the reality sufficiently closely. One of the main
reasons is that in practice (i.e., up to a few most elementary
quantum systems), the {\em physics} behind the interaction often
proves complicated: relativistic or nonlocal or nonlinear.

In our present text we repeatedly pointed out that our present
linearized and perturbed square-well model is certainly interesting
{\it per se}. It might fulfill the role of an interesting effective
model in physics. Still, its methodical relevance is related to the
nonlinear setting of paper I, potentially useful in the context of
the study of quantum systems described by certain prohibitively
complicated conventional linear Schr\"{o}dinger equations
(\ref{seor}).

One of the fairly instructive testing grounds of the efficiency of
the suppression of the technical complications via non-linearization
may be found in classical optics where certain deeply relevant
dynamical effects may be very efficiently described via a transition
 $
 V(\vec{x}) \to V(\psi(\vec{x}))
 $
to a suitable state-dependent interaction term. With one of the
least complicated tentative choices of $V(\psi(\vec{x})) \sim
\psi^*(\vec{x})\psi(\vec{x})$ one arrives at the highly popular toy
model called ``non-linear Schr\"{o}dinger equation''
\cite{nlse,nlseb,nlsec}. In this context, we tried to find a formal
encouragement and support also for the logarithmic self-interaction
in our present letter.

In practice, typically, the strictly linear theory only remains
friendly and feasible for the most elementary systems like hydrogen
atoms, etc. Moreover, even in the phenomenology based on the linear
equations one of the key roles is played by the educated guess or
knowledge of the relevant  {\em dynamical input information} about
the {\em linear} interaction. Thus, one may conclude that in this
language the transition to the {\em effective} nonlinear models
(including (\ref{nelie})) does not look drastic or counterintuitive
to a physicist.

A number of supportive phenomenological arguments may be found in
paper I or, e.g., in the recent remark \cite{vjes} on the effective
nonlinear logarithmic Schr\"{o}dinger equations
 \be
  i \partial_t \psi(\vec{x},t)
  =(- \Delta +V_{LSE})\, \psi(\vec{x},t)\,,
  \ \ \
  V_{LSE}=- b\, \ln{|\psi(\vec{x},t)|^2}
    ,
  \label{lnse}
  \ee
where their relevance in the phenomenology of quantum liquids has
been emphasized. Alas, the situation may become perceivably more
complicated in mathematics. Multiple challenges emerge there. In
their light, our present letter may be perceived as constructive
commentary on these complications, i.e., as a contribution to a
future completion of the formalism of practical quantum mechanics
\cite{Fluegge}.

\subsection*{Acknowledgements}

The project was supported by GA\v{C}R Grant Nr. 16-22945S. Iveta
Semor\'{a}dov\'{a} was also supported by the CTU grant Nr.
SGS16/239/OHK4/3T/14.


%
%
%
%

\end{document}